\documentclass[a4paper]{easychair}
\usepackage[english]{babel}
\usepackage[T1]{fontenc}
\usepackage{fancyhdr}
\usepackage{algorithmicx}
\usepackage{algorithm,algpseudocode}
\usepackage{float}
\usepackage{amssymb}
\usepackage{amsmath}
\usepackage{amsthm}
\usepackage{amsfonts}
\usepackage{mathrsfs}
\usepackage{graphicx}
\usepackage{subfigure}
\usepackage[shortlabels]{enumitem}
\usepackage{multicol}
\usepackage[toc]{appendix}
\usepackage{midpage}
\usepackage{geometry}
\usepackage[font=small]{caption}
\usepackage[utf8]{inputenc}
\usepackage{textcomp}
\usepackage{xcolor}
\usepackage{framed}
\usepackage{tabularx}
\usepackage{tikz}
\usepackage{tikz-cd} 
\usepackage{booktabs}
\usepackage{multirow}
\usepackage{csquotes}
\usepackage{varioref}
\usepackage{hyperref}
\usepackage[nameinlink, capitalise, noabbrev]{cleveref}

\definecolor{lime}{HTML}{A6CE39}
\DeclareRobustCommand{\orcidicon}{%
	\begin{tikzpicture}
	\draw[lime, fill=lime] (0,0) 
	circle [radius=0.16] 
	node[white] {{\fontfamily{qag}\selectfont \tiny ID}};
	\draw[white, fill=white] (-0.0625,0.095) 
	circle [radius=0.007];
	\end{tikzpicture}
	\hspace{-2mm}
}

\foreach \x in {A, ..., Z}{%
	\expandafter\xdef\csname orcid\x\endcsname{\noexpand\href{https://orcid.org/\csname orcidauthor\x\endcsname}{\noexpand\orcidicon}}
}

\makeatletter
\def\BState{\State\hskip-\ALG@thistlm}
\makeatother

\newcommand{\mult}[1]{\mathtt{#1}}

\theoremstyle{plain}

\crefname{thm}{theorem}{theorems}
\Crefname{thm}{Theorem}{Theorems}

\crefname{cor}{corollary}{corollaries}
\Crefname{lem}{Corollary}{Corollaries}

\crefname{prop}{proposition}{propositions}
\Crefname{prop}{Proposition}{Propositions}

\crefname{lem}{lemma}{lemmas}
\Crefname{lem}{Lemma}{Lemmas}

\theoremstyle{definition} 

\crefname{defn}{definition}{definitions}
\Crefname{defn}{Definition}{Definitions}
\theoremstyle{remark}
\crefname{remark}{remark}{remarks}
\Crefname{remark}{Remark}{Remarks}

\tikzstyle{mybox} = [draw=black, very thick, rectangle, rounded corners, inner ysep=5pt, inner xsep=5pt]

\newcounter{protocol}
\newenvironment{protocol}[1]
{\par\addvspace{\topsep}
	\hrule
	\vspace{5pt}
	\noindent\refstepcounter{protocol}\textbf{Protocol \theprotocol} #1 \\
	\vspace*{-5pt}
	\hrule
	\small}
{
	\hrule
	\vspace*{5pt}
	\par\addvspace{\topsep}}
\Crefname{protocol}{Protocol}{Protocols}

\newtheorem{prob}{Problem}[section]{\bfseries}{\itshape}
\crefname{prob}{problem}{problems}
\Crefname{prob}{Problem}{Problems}

\begin{document}

\title{Cob: a consensus layer enabling sustainable sharding-based consensus protocols}
\newcommand{\orcidauthorA}{0000-0002-3872-7251}
\newcommand{\orcidauthorB}{0000-0002-8739-3091}
\newcommand{\orcidauthorC}{0000-0002-5159-7252}

\author{
    Andrea Flamini\thanks{\email{andrea.flamini.1995@gmail.com}\hspace*{\fill}\orcidA{}
     https://orcid.org/0000-0002-3872-7251}
     \and
    Riccardo Longo\thanks{\email{riccardolongomath@gmail.com} \hspace*{\fill}\orcidB{}
     https://orcid.org/0000-0002-8739-3091}
    \and
    Alessio Meneghetti\thanks{\email{alessio.meneghetti@unitn.it} \hspace*{\fill}\orcidC{}
     https://orcid.org/0000-0002-5159-7252}
}
\institute{Department of Mathematics, University of Trento\\ via Sommarive, 14 - 38123 Povo (Trento), Italy }

\titlerunning{Cob: Consensus for Sustainable Sharding-Based Protocols}
\authorrunning{A. Flamini, R. Longo, A. Meneghetti}

\maketitle

\begin{abstract}In this paper we explore a context of application of Cob, a recently introduced Byzantine Fault Tolerant consensus protocol.

Cob proves to be a leaderless consensus protocol which carries out the consensus process in parallel on each component of a list of events to be observed and recorded.

We show how Cob can be used to define a consensus layer for scalable and sustainable blockchains.
This layer is used to design consensus protocols based on \emph{sharding} as a mean to achieve scalability, and on the \emph{fragmentation of time in time-slots} (which get assigned to nodes that are instructed to create new blocks) as a mean to reduce the amount of computation and communication necessary for the maintenance of the distributed ledger.

We explain why Cob is a viable candidate to implement such consensus layer through the introduction of an auxiliary blockchain that we name Synchronization Chain.
\end{abstract}

\section{Introduction}
\label{intro}
A blockchain is a distributed ledger which allows a network of nodes to record transactions in a trusted and immutable manner.
The network is generally composed of independent parties which cooperate to the maintenance of the ledger without the influence of a central authority. Clearly, since a blockchain is a distributed system, one of the most important problems is the consensus.
Every blockchain is provided with a consensus algorithm which allows the network of nodes to agree on the information to record into the ledger, even in presence of malicious or faulty nodes.
Since consensus protocols can be quite intensive under the computational or communication point of view (e.g. proof of work or Byzantine fault tolerant protocols, respectively), the number of transaction that can be recorded in the ledger per second in average is low if compared with the centralized counterparts.
Therefore, the same negative comparison can be done regarding the costs applied to the users in terms of fees (e.g Bitcoin \cite{nakamoto2008bitcoin} vs Visa).
This causes blockchain platforms to be incapable to grow and manage an increasing number of requests, a property which is referred to as \emph{scalability}.
In particular, we say that a platform \emph{scales} if it can easily adapt to changes in the number of users that decide to join in, as well as in the number of transaction requests that such users perform.

\subsection{Preliminaries}
\label{prel}

We now introduce two techniques which are adopted to solve the problem of intensity of computations/communications, and the problem of scalability.

\subsubsection{Reduction of computations or communications}

Some consensus protocols, such as EOS \cite{EOS2017}, Quadrans \cite{battagliola2021quadrans} or Takamaka \cite{takamaka} divide the epochs in smaller time units which we call time-slots.
Before the new epoch begins, a redistribution mechanism assigns each time-slot to a node in the network.
These nodes are in charge of the creation of a block and if one of them does not manage to broadcast its block in time (i.e. before the end of the time-slot), then its block gets discarded by the network.
In this approach, the network first reaches a consensus on the way the time-slots must be redistributed and, after that, only the node in charge during a time-slot can produce and advertise a new block.
This drastically reduces the computational consumption derived by the classical protocols based on proof of work such as Bitcoin \cite{nakamoto2008bitcoin} which requires that the nodes execute intensive computations without any break.
Similarly, this approach reduces the burden of communications between nodes given by Byzantine fault tolerant (BFT) protocols such as Algorand \cite{chen2019algorand} or HoneyBadgerBFT \cite{miller2016honey}.

However, the subdivision of an epoch in preassigned time-slots, although it brings several benefits in term of energetic efficiency and platform stability, it also brings one important issue: every node in the network must have access to a common clock which specifies the beginning and the end of a time-slot, an essential tool to determine whether the node in charge created the block in time or not.
The problem of accessing a common clock could be avoided by using same speed clocks and an event which triggers the end of a time-slot and the beginning of a new one\footnote{Some consensus protocols such as Algorand \cite{chen2019algorand} assume that the nodes in the network are provided with same speed clocks and each node resets its clock every time a certificate for the new block is received. Since Algorand assumes that the certificate propagation time is below a parameter $\lambda$, the delay between two nodes is upper-bounded by $\lambda$.}.
But, even with a common clock, there would still be the problem that a message does not reach every node in the network in the very same instant of time.
Therefore, a block may be received in time by some nodes, and the same block may be received late by some other nodes if it is broadcast near the end of the time-slot.
In fact, in a distributed system where the messages are broadcast via gossip, the time in which a node $n$ receives a message does not provide very precise information about the time the other nodes have seen such message.
Let $\lambda$ be the message propagation time, assuming a common clock exists and $T$ is the time when $n$ has received the block, then, if the communication happens via gossip (which is typical in the context of permissionless blockchain networks), the other nodes will receive (or have received) the block in the time interval $[T-\lambda,T+\lambda]$.
The node $n$ does not know much more than this.

This observation opens the door to a series of vulnerabilities of the system which may allow an attacker to discard blocks legitimately created and diffused by an honest node.
In fact, if there is not a third party who certifies the legitimate creation and diffusion of a block, the only feedback the network receives regarding the timing of diffusion consists only of the opinion of the nodes in charge in the following time-slots.
\bigskip

In \Cref{applic} we will explain how Cob, a novel BFT consensus protocol, can substitute the third party which attests which blocks have been legitimately created in time by the right node.
This protocol is executed by the network of nodes and is a \emph{leaderless consensus protocol}, therefore the only way for a block to get certified and accepted, is to be received in time by a great majority of the nodes in the network.
This is exactly what we expect to happen when a node behaves honestly, and makes it impossible for an attacker to pretend that such block was not diffused in time.

Moreover, the publication of the certificates that attest that a block have been created in time can be used as the triggering event which officially ends the current time-slot and starts the new one.
This approach to the declaration of the beginning of a new period of time is similar to the one adopted by Algorand \cite{chen2019algorand} to declare the beginning of the next round.

A legitimate question might be why not to use a BFT protocol to reach consensus directly on the new block instead of reaching consensus on a certificate which proves the legitimacy of the newly created block.
The quick answer to this question is that reaching consensus on a certificate will let us prove the legitimacy of multiple blocks simultaneously, and delegate expensive and/or difficult checks on the individual blocks, as it will become more clear in the following sections.

\subsubsection{Improving the scalability}
\label{sharding}
In order to solve the scalability issues of blockchain platforms, many approaches have been proposed over the years.
Some of them are the \emph{block size increase}, the use of \emph{off-chain state channels}, \emph{segregated witness (SegWit)}\cite{lombrozo2015segregated}, the use of \emph{directed acyclic graphs} as in \cite{sompolinsky2016spectre} and \emph{sharding}.
Among these, sharding seems to be the most promising \cite{liu2021building,wang2019sok}, a description of the main platforms adopting this technique is presented in \cite{meneghetti2019survey}.

The term sharding comes from database management, where it identifies a particular type of database partitioning, that consists in dividing large databases into smaller parts, called shards.
Shards are more manageable in terms of server hosting and other aspects of database maintenance, and allow to have faster query time by diversifying the responsibility of a database structure.
Similarly, when we talk about sharding in the context of blockchain platform design, we refer to an architecture which divides the ``usual'' chain of blocks into multiple chains called shards, which are managed in parallel by different groups of nodes.
This improves throughput, since many transactions can be simultaneously validated, allowing blockchains to effectively scale for a huge number of users.
Although sharding is promising, it faces many challenges that the community must efficiently and securely solve.
For example, one should note that, if a network is divided in shards, for an adversary it is potentially easier to take control of a single shard compared with the whole network.
In fact, its impact in terms of ratio of nodes it controls grows linearly with the number of shards the network adopts to record transactions.
Another challenge the protocol designers must deal with is the inter-shard communication: nodes working on different shards might be in possess of or access to different data sources.
Therefore, the protocol designer must assure that transactions elaborated by different shards are consistent despite the fragmentation of the transaction insertion process.

Since sharding aims to maximise the scalability of a platform depending on the underlying network of nodes (with great focus on preserving the security requirements), and the network conditions can suddenly evolve\footnote{The network evolution refers to: a) new nodes who decide to join the network or nodes that decide to leave it, b) nodes who decide to actively partake to the consensus process and others which decide to be passive and only have access to the information recorded into the ledger, c) nodes which become faulty over time (or, more importantly, an attacker corrupts some of them whenever it believes it is profitable).}, it is good practice to regularly update the configuration of the system (i.e. the actors and parameters involved).
Each period of time in which the system configuration is updated is referred to as \emph{epoch}.

In order to better comprehend what sharding is, we briefly describe some of the main components of a consensus protocol for a blockchain implementing sharding.
We refer to the surveys \cite{wang2019sok,liu2021building} for more details about blockchain sharding.
\begin{enumerate}
    \item \emph{Identity establishment and shard formation}: this process aims to identify the single nodes who take part to the protocol execution and (randomly) assign them to a specific shard.
    This process should prevent Sybil attacks from being successfully performed by malicious entities who manage to create multiple identities.
    
    \item \emph{Intra-shard consensus} Each node within a shard must execute a consensus protocol to reach agreement on the transactions to be recorded in the fragment of ledger which is under that shard's control.
    Here, we make a distinction between two possible approaches to intra-shard consensus: \emph{weak} and \emph{strong consistency}.
    According to the definition in \cite{liu2021building}, weak (or eventual) consistency means that different nodes might end up having different views of a blockchain, which leads to forks, therefore a certain number of blocks at the end of the blockchain need to be truncated to obtain stable transactions.
    Contrarily, strong consistency means that after the generation of a valid block, every non-faulty node shares the same view and therefore no forks can happen.
    
    \item \emph{Cross-shard transaction processing}: it is essential that the transactions processed by the shards are consistent not only within the shard they belong to, but also across the whole system.
    Therefore, for cross-shard transactions, which are transactions which involve information processed by more than a shard, a network must adopt some mechanism which allows synchronization and reconciliation of transactions processed by different groups.
    
    \item \emph{Epoch reconfiguration}: in order to guarantee the security of the shards, they may need to be reconfigured, requiring both reconfiguration rules (to let the platform respond to the network evolution) and possibly a randomness source.
\end{enumerate}

In \Cref{applic} we will explain why Cob can be adopted to deal with the epoch reconfiguration and can contribute to the cross-shard transaction processing in a sharding-based consensus protocol.

\subsection{Contribution}

Cob \cite{Cob} is a novel BFT protocol (i.e. a strong consistency protocol) which is an evolution of the MBA protocol \cite{flamini2022multidimensional}.
MBA is defined for complete synchronous networks, therefore it can be executed only by small networks with few dozens of nodes, while Cob can be executed by incomplete gossiping networks that can have millions of nodes, a property that makes it suitable for networks of permissionless blockchains.
The aim of Cob is to allow a network of nodes to reach consensus on a list of time-stamps of events that are expected to happen in a time interval, we will explain why it can be adopted in the design of consensus protocols for blockchains implementing sharding.

In this context, Cob can be used to let the network make multiple decisions simultaneously, for example decisions about which shards have correctly performed their job, or decisions about the evaluation of parameters which characterize the protocol epochs on the basis of the network conditions.
We will emphasize the  advantages that Cob brings to the organization of the workload that must be executed by the shards and show some performance evaluations regarding a Cob execution under the framework we will describe.

\paragraph{Outline}

In \Cref{protocol} we provide a high-level but comprehensive description of Cob, which is fully presented in \cite{Cob}.
In particular we underline the properties that Cob satisfies and the assumptions it relies on.
In \Cref{highLevel} we describe its workflow and building blocks, namely the Multidimensional Graded Consensus and Multidimensional Binary Byzantine Agreement \cite{flamini2022multidimensional,Cob}.

In \Cref{applic} we describe how Cob can be used to create a framework for the design of sustainable and scalable blockchain platforms.
Scalability is obtained by using the sharding technique, sustainability is obtained by dividing time in time-slots, during which some prescribed nodes are expected to create blocks of processed transactions. In \Cref{synchrochain} we propose a solution to put in practice the principles previously presented.

Finally, in \Cref{performance} we compare the performance of Cob and Algorand as consensus protocols for the solution presented in \Cref{synchrochain}. The comparison is centered into quantifying the amount of data broadcast in the network by the nodes.

\section{The consensus protocol Cob}
\label{protocol}
In this section we provide a high-level description of \emph{Cob}, a novel consensus protocol which efficiently solves the following problem:

\begin{prob}\label{general_problem}
    Given a set of events which a network of nodes can observe, how can the nodes reach consensus on some relevant information about such events?
\end{prob}

The problem above is discussed in \cite{flamini2022multidimensional, Cob} by considering the presence of malicious nodes in the network, and this leads to the identification of two properties that such a consensus protocol must satisfy in order to maximise the amount of agreed-upon meaningful data.

\begin{enumerate}
    \item The consensus protocol must be \emph{leaderless}, which means that there is no single node proposing a protocol output and the other nodes decide whether to accept it or not, but rather several nodes must collectively construct the output list.
    The reason behind this is that a leader, if honest, would propose a list of relevant information which is heavily influenced by its own point of view (which in some cases might lead to incorrect decisions), and if the leader is malicious, it may easily perform censorship attacks refusing to include some information in the list, or deliberately include invalid information.
    In both cases, if the network does not agree even with one component proposed by the leader, it will reject the leader's proposal, and this process is repeated until a leader proposes a list which gets accepted by a majority of the network.
    Note that this might not even happen, in fact, if there is a wide disagreement among the nodes about one or more components, there might not exist a list which is accepted by such majority.

    \item The consensus process must be carried out in \emph{parallel} and independently on each component of the list, so that disagreement on a single component does not affect the consensus on the others.
    In this regard, if a specific component can not be agreed upon on any meaningful value due to a wide disagreement among the nodes of the network, then the network must be able to identify this network condition and manage to reach consensus on a default value that we identify with $\bot$.
\end{enumerate} 

In \cite{flamini2022multidimensional} there is a simple example which helps to understand why these two properties have such a great impact on the way consensus is achieved.
In \cite{flamini2022multidimensional} it is also presented a predecessor of Cob as a solution to \Cref{general_problem} for a relatively small network of a fixed number $n$ of nodes, under some strong communication assumptions, considering an attacker that controls less than $\frac{1}{3}$ of the nodes. In particular it is assumed a strongly-synchronous communication model and a complete network, where every node could instantaneously send a message to every other node.
These assumptions are not practical and dramatically reduce the number of application contexts.
In fact, under the \emph{complete and synchronous network} (CS network) model, the communications between nodes happen instantaneously, via direct channels, every time a common clock (i.e. shared by all the network participants) ticks the beginning of a new protocol step.
Since it is assumed that the network is complete, it is essential, for sake of efficiency, that the network is composed of a small number of nodes (which does not exceed the hundreds).
The evolution of Cob presented in \cite{Cob} overcomes these shortages, and can be adopted by wide networks of nodes (even with millions of participants), making it way more practical and its adoption suitable for the network of permissionless blockchains.
From now on, when we refer to the protocol Cob we refer to the second definition \cite{Cob}, unless explicitly specified.

\subsection{Cob: network and communication assumptions}
Since in complete networks the number of messages exchanged through the network
grows exponentially with the number of network participants, for practical
applications it is more convenient to consider a network model which differs from the CS network, such
as the \emph{Asynchronous Gossiping Network} (AG network)\footnote{ Algorand describes the environment in which it is defined as asynchronous. This is because the communications between nodes happen via gossip and the protocol steps, which for a single user are non-overlapping time intervals, for different users may overlap due to lack of clock synchronization.
However, since Algorand assumes that there is a predetermined upper-bound to the time required by a message to reach (almost) every node, and therefore there is an upper-bound to the delay between different nodes, Algorand can not be considered an asynchronous protocol.} presented by Micali
in Algorand~\cite{chen2019algorand}.

In this model messages are broadcast in the network in a gossiping fashion: a procedure characteristic of peer-to-peer communications where messages pass from one node to its neighbours and so on until they reach every node.
In gossiping networks the network relies on each member to pass messages along to its neighbours, therefore it is reasonable to envisage the network as an incomplete, connected and non-directed graph.
We assume that a message sent by an honest node reaches every honest node within a time limit that depends on the size of the message itself.
Since malicious nodes can behave arbitrarily, the previous assumption means that they cannot be cut vertices in the network graph, that is the graph remains connected even without the edges connected to malicious nodes.
We will also require that the ratio of malicious or faulty nodes is less than $\frac{1}{3}$.

\subsubsection{Timing assumptions}
In an AG network there does not exist a common clock (as in the case of CS network), but it is assumed that all network participants are provided with \emph{Same-Speed Clocks} \cite{chen2019algorand}.
In other words, it is assumed that each network participant has its own clock and that the clocks all have the same speed, even if they are not synchronized in any way.
However, it is assumed that there is an upper-bound $\lambda$ on the time required by a node to diffuse in the network a "short" message.
Therefore, this assumption implies that the non-synchronized clocks can reach a sort of synchronization in the following way: suppose that a node communicates to the other nodes, via a short message $M$, the beginning of a new protocol execution.
This node will immediately reset its private clock to 0, as the broadcast of $M$ begins, and the other nodes will do the same once they are reached by $M$.
Since the message $M$ reaches every node in the network within time $\lambda$, every node will reset to 0 their own private clock in the absolute time interval [0,$\lambda$] (where 0 is the absolute time when the first node broadcasts $M$), causing the delay between different nodes to be upper-bounded by the parameter $\lambda$.
Afterwards the time discrepancies do not vary because of the same-speed nature of the clocks.

We have explained how an AG network addresses the goal of designing a practical model which, on one hand it does not require a node to send a message to every single node every time it wants to share some information with the whole network, but on the other hand it forces the nodes to maintain their private same speed clocks slightly asynchronous (with the delay which is upper-bounded by a constant value $\lambda$).

\subsubsection{Sortition mechanism}
Another relevant aspect regarding the design of Cob is the following: since the nodes in the network which adopts Cob as consensus protocol can be as wide as needed, it is essential that not every node in the network broadcasts a message at the end of every step.
This would clearly cause a network overload. In order to address this problem, Cob uses a \emph{sortition mechanism} which instructs some nodes to be active during a given protocol step (i.e. to broadcast a message) while assigning to the other nodes a passive role (i.e. just collect and help broadcast the messages of the players).
In order to better clear up this distinction, from now on, in a specific step, we will call \emph{players} only the nodes selected to be active and broadcast their message, while a generic node of the network will be referred to as a \emph{user}.

\subsection{High level description of Cob}
\label{highLevel}

We now provide a high level description of the protocol.
We first describe two protocols which are the building blocks of Cob and we explain how they achieve the core properties mentioned at the beginning of this section: the fact that Cob is leaderless, and that the consensus process is carried out in parallel on the components of the list.
The first component is the \emph{Multidimensional Graded Consensus (MGC)} and the second one is the \emph{Multidimensional Binary Byzantine Agreement (MBBA)}.
Both protocols are an extension to the multidimensional case of protocols presented by Micali and adopted in Algorand \cite{chen2019algorand}, namely the Graded Consensus protocol \cite{feldman1997optimal} and the Binary Byzantine Agreement protocol \cite{micali2016byzantine}.

\subsubsection{Cob's building blocks}

The MGC is a 3 steps protocol which starts once the network observes the events they want to time-stamp and requires the players (i.e. the users who are elected to be active in a given step) of the first step to broadcast their list of observed values created during the observation phase.
Once the 3 steps are executed, each node $n$ in the network privately builds a list $\mathbf{v}^{(n)}$ of relevant information about the observed events (note that the lists can eventually be different for different nodes).
Together with the list of relevant information, each node $n$ computes a grade $g_i^{(n)}\in \{0,1,2\}$ associated to each component $i$ of the list, which represents the confidence that such value is well spread around the network, according to the information received.
In particular, a grade of 0 represents a high disagreement in the network; 1 represents a state of uncertainty given by an intermediate number of messages advertising the corresponding value; 2 guarantees the node which computed the grade that every honest nodes has recorded the same value in that component.

Distinct nodes might have saved different lists of relevant information, and they might have also recorded different grades, based on the messages received in the steps of MGC Protocol.
However, it is proven that, for each component, the difference between the grades of honest nodes is $\mid g_i^{(n)}-g_i^{(m)} \mid\le 1$ and it is also proven that, if $g_i^{(n)}\ge 1$ for each honest node, then the relevant information recorded in the $i$-th component is the same for every honest node.
This implies that if some honest node sets $g_i^{(n)}=2$ then the relevant information saved by the nodes in the $i$-th component is the same for every node. This is a remarkable information, but we recall that the protocol is executed by nodes that do not trust each other. Therefore, it is necessary to find a way to let the nodes who are certain that a relevant information is shared by all the honest nodes convince the nodes who are not certain about it.
For this purpose the network executes the protocol MBBA: a 3 steps loop which allows the nodes in the network to reach agreement on a list of bits with the same dimension of $\mathbf{v}^{(n)}$.
The scope of this protocol execution is to detect the components of the list of relevant information $\mathbf{v}^{(n)}$ which are the same for every honest node.
In particular, after the MGC protocol execution, each node will build a list of bits setting each component to: 0 if it is assured that all the nodes are in agreement on that specific component (i.e. the associated grade is 2), 1 otherwise. Now the network is ready to perform the MBBA protocol, since every node has its own private initial list of bits. In \cite{flamini2022multidimensional,Cob} it is proven that the MBBA is a Byzantine Agreement protocol, which allows a network of nodes provided with an initial list of bits to reach consensus on a shared list of bits $\mathbf{b}$.
In the context of Cob, the private initial list will be computed from the output of MGC, but at the end of the MBBA execution, the nodes will reach consensus on $\mathbf{b}$, which allows every node $n$ to compute the list of relevant information (the output of Cob) on which the honest nodes can be in agreement.
This list $\mathbf{v}$ is built in the following way: for every honest node $n$ $\mathbf{v}_i=\mathbf{v}^{(n)}_i$ if $\mathbf{b}_i=0$ otherwise $\mathbf{v}_i=\bot$, which means that such component is left blank.

In \cite{Cob,flamini2022multidimensional} it is proven that Combining MGC with MBBA it is possible to solve \Cref{general_problem}.
We omit the formal definition, and just summarize the main protocol steps of Cob using the building blocks previously described.
For a detailed description of the protocol we refer to~\cite{Cob}.

\begin{protocol}{\emph{Cob}}
\begin{itemize}
    \item \texttt{Observation Phase}
For every user $u$ in the network:
\begin{itemize}
    \item $u$ observes the events $\mathbf{E}=(E_1,\dots,E_m)$ that must be time-stamped;
    \item $u$ locally records the observed values $\mathbf{o^{(u)}}=(o^{(u)}_1,\dots,o^{(u)}_m)$.
\end{itemize}

\item \texttt{Multidimensional Graded Consensus} 
For every user $u$ in the network:
\begin{itemize}
    \item $u$ starts the execution of MGC with input the list $\mathbf{o^{(u)}}$;
    \item $u$ takes part actively to a step of MGC only if it is elected as a player via the random sortition mechanism adopted by Cob;
    \item output 1 of MGC: $u$ locally saves the list of values $\mathbf{\Theta^{u}}=(\Theta^{u}_1,\dots,\Theta^{u}_m)$, given by MGC; 
    \item output 2 of MGC: from the list of grades $\mathbf{g^{u}}=(g^{u}_1,\dots,g^{u}_m)$ given by MGC, $u$ obtains a list of bits $\mathbf{v^{u,0}}=(v^{u,0}_1,\dots,v^{u,0}_m)$, where $\forall i\in\{1,\dots,m\}$, ${v^{u,0}_i=0 \iff g^{u}_i=2}$, and $v^{u,0}_i=1$ otherwise.
\end{itemize}

\item \texttt{Multidimensional Binary Byzantine Agreement} 
For every user $u$ in the network:
\begin{itemize}
    \item $u$ starts the execution of MBBA with input the list of bits $\mathbf{v^{u,0}}$;
    \item $u$ takes part actively to a step of MBBA only if it is elected as a player;
    \item output of MBBA: $u$ builds a certificate for $\mathbf{v^{u}}=(v^{u}_1,\dots,v^{u}_m)=\mathbf{v}=(v_1,\dots,v_m)$, which is the same for every honest user in the network.\footnote{What the network is actually doing during the MBBA execution is identifying the components of the vectors $\mathbf{\Theta^{u}}$ which are the same for every honest user $u$.
    In particular if agreement on a component $c$ of $\mathbf{v}$ (the list of bits) is reached on 0, i.e. $v_c=0$, then this means that the honest users share the same value $\Theta^{u}_c$ and they will preserve it, otherwise, if agreement has been achieved on 1, i.e. $v_c=1$, this means that the network could not be convinced that the honest nodes share the same value $\Theta^{u}_c$.}
\end{itemize}

\item \texttt{Cob Output Determination}
Being $\mathbf{v}$ the output of MBBA and $\mathbf{\Theta^u}$ the first output of MGC computed by the user $u$, $u$ computes the output of Cob $\mult{out^{u}}=(\bar\Theta^{u}_1,\dots,\bar\Theta^{u}_m)$, where $\forall i\in \{1,\dots,m\}$, $\bar\Theta^{u}_i=\Theta^{u}_i$ if $v_i=0$ and $\bar\Theta^{u}_i=\bot$ if $v_i=1$. \footnote{It is proven that for each pair of honest users $u_1, u_2$, $\mult{out^{u_1}}=\mult{out^{u_2}}$ holds.}
\end{itemize}
\end{protocol}

We underline the fact that the way MBBA is used in Cob is the same way BBA is used in Algorand: the goal is to decide whether to reject or accept a candidate piece of information (for Algorand a block, for Cob an observed value or a relevant information about a give event).
However, MGC is used in a very different way: while GC in Algorand is used to determine the leader of a given protocol run, MGC in Cob is used to collect the opinion of several nodes advertising the list of values they have observed\footnote{ Algorand is a leader-based consensus protocol for a blockchain used to exchange cryptocurrency, therefore it tolerates that some blocks may be created by malicious nodes and contain no transactions.
In fact, in many applications it is not necessary that a transaction request is immediately included in the newly created block, what is essential is that eventually an honest node will create a block which includes the pending transaction request.}.

\section{Applicability of Cob}
\label{applic}
In this section we will explain how Cob can be used to regulate a scalable and sustainable blockchain.
As we explained in \Cref{prel}, an approach which can be used to reduce the amount of computation or communication necessary to the maintenance of a blockchain is the subdivision of time into preassigned time-slots.
This concept clearly can be combined to sharding  in order to increase the scalability of a blockchain platform.
What we obtain is a sharding-based blockchain in which for each shard the network assigns a time-slot to a specific node.
Therefore, if the number of shards in a given epoch is $m$, then there must be $m$ nodes, one for each shard, who are expected to publish a block within the end of the time-slot.
But let us proceed step by step in the description of how Cob can be used to achieve this protocol structure.

\subsection{Cob and time fragmentation}
\label{time_frag}
First, we explain how Cob can be adapted to the certification of block creation.
As we have mentioned in \Cref{protocol}, Cob is a leaderless consensus protocol which has been designed to let a network of nodes reach consensus on the description of a set of events which are expected to happen in a time interval.
If we can consider a standard blockchain (a single shard) which is maintained with the use of preassigned time-slots, then there is a single event that the network observes during every time-slot: the creation of a block performed by the node in charge.
Note that the fact that Cob carries out the consensus in parallel on each component of the list of events observed is not relevant now (since we are considering only one event), but it will become relevant in \Cref{Cobshard}.

If the following statements hold:
\begin{enumerate}
\item nodes are provided with 
same speed clocks;
\item it is possible to upper-bound the diffusion time of a message of fixed weight;
\item  the nodes agree on $t$, the duration of a time-slot;
\item the network agrees on a list $L$ which assigns each time-slot o a given node;
\end{enumerate}
then we can describe a protocol which manages the definition of time-slots and guarantees that an attacker can not pretend it has received late a legitimately created block.

In \Cref{Cobshard} we will explain how to obtain the last two items of the list, namely the duration $t$ of the time-slots and the list $L$; for what concerns the first two items, they are commonly adopted assumption in distributed protocol definitions.
Assuming that each network member is in possess of the information above, the protocol could work in the following way:
\begin{protocol}{Time-slot}\label{protocol-time-slot}
\begin{itemize}
    \item \texttt{Synchronization setup}: the network executes an instance of Cob to decide when to start; as soon as the network creates a certificate for the message \texttt{start}, the actual protocol can begin. 
    
    Since the upper-bound for the diffusion time of the certificate is $\lambda$, the delay between any two honest nodes is less than $\lambda$.
    
    
    \item \texttt{Block creation}: the node in charge, according to the list $L$ builds a block of transactions and before time $t-2\lambda$ broadcasts this block.
    
    
    \item \texttt{Timing evaluation}: the nodes of the network start executing Cob when their own private clock signs time $t$, and try to reach consensus on the digest of the newly created block.\footnote{Note that if the creator of the block has broadcast it before $t-2\lambda$ (according to its own time reference), then every node has received the block within time $t$ (again, according to their own time reference), since $\lambda$ upper-bounds the time for the message diffusion, and also the delay between two nodes.} 
    
    
    \item \texttt{Certificate creation and start of a new time-slot}: each player marks the end of the current time-slot and the beginning of the new one as soon at the reception of a certificate for the newly created block (produced by the network via Cob).
    The nodes reset their same-speed clocks (the delay is given by the diffusion time of a certificate, which is again $\lambda$).
    
    Return to \texttt{Block creation}.
\end{itemize}

\end{protocol}

In \Cref{protocol-time-slot} it is shown that, assuming the existence of a list $L$, which assigns each time-slot to a node, this iterative protocol guarantees that if the right node creates and broadcasts a block in time, the network can certificate the correctness of the creation process. 

Up to now, the network has not evaluated the transactions included into the block, however, this can be done right after timing verification.
Assuming that $L$, the consensus protocol, and the semantic rules which define which transactions can be included into the ledger are public knowledge, the only aspect that can bring the nodes to disagreement is whether the legitimate node has created its block in time.
The disagreement may be caused by the delay between nodes and the time of diffusion of messages.
Once this information is agreed upon through Cob, every honest node will be able to determine if the transactions are invalid (therefore the block will not be taken into account) or the block can be preserved.

\subsection{Cob for time-slot assignation and sharding consensus}
\label{Cobshard}
As emphasized in \Cref{sharding}, a key concept in the design of blockchains implementing sharding is the \emph{epoch}: a time interval in which the system configuration (i.e. protocol parameters and nodes partaking the consensus process) is fixed \cite{liu2021building}.
Once the epoch ends, the actors executing the consensus protocol may be substituted and, if the network has evolved, the protocol parameters can be updated accordingly.

While explaining how Cob can be useful in the implementation of architectures based on sharding, we will follow the guidelines described in \Cref{time_frag}, maintaining the division of time in preassigned time-slots for each shard.
We remind that our goal is to propose a consensus layer which may help in the creation of sustainable and scalable blockchain platforms.
The first adjustment that must be done regards the list $L$ which deals with the time-slot assignation, which must cover the time-slots of an epoch and then must be updated for the following one.
Since in the sharding case there is more than one chain of blocks, the list $L$, together with the time-slot, must specify the shard on which a node must append its block.

Cob can be very useful in the definition of a sharding based architecture mostly by periodically making a freeze frame representing the network status and defining the next epoch configuration.
Since Cob is a strong consistency protocol, once the frame describing the network status is published, the network will consider those information final, and act accordingly.
For instance, Cob can be used to let the network determine the protocol parameters of an epoch on the basis of the information previously broadcast by the nodes.
Examples of protocol parameters that must be agreed upon to define an epoch are the following:
\begin{enumerate}
    \item the number of shards;
    \item the number of time slots;
    \item the duration of the time-slots;
    \item the list $L$ of nodes in charge of the creation of a block in a given time-slot and a given shard.
\end{enumerate}

Note that these parameters must be determined according to the information observed during the previous epochs, and the consensus protocol must make explicit some deterministic rules to let the network easily reach consensus on the values.
In fact we recall that consensus can be reached if there exist, at the beginning of the protocol, a sufficiently large agreement on the values proposed by the single nodes: then the consensus protocol makes this agreement explicit, reliable and final. 
Therefore, the evaluation of the epoch parameters can be seen as a description of events observed during the previous epochs.
For example, the number of shards active during an epoch could depend on the number of nodes that apply for becoming block creators in the following epoch: the higher the number of candidates, the higher the number of shards.
This means that each protocol parameter can be seen as a description of events observed in a given time interval, therefore they can be determined executing an instance of Cob.

It is necessary to clarify the fact that some parameters depend on other parameters, for example the list $L$ depends on the number of shards and the number of time-slots.
In this case it is sufficient to validate the list $L$ to consequently implicitly fix the number of shards and the number of time-slots.

\subsection{The Synchronization Chain based on Cob}
\label{synchrochain}
Now that we have explained how Cob can be useful in the creation of a sustainable and scalable blockchain, the next question is: how can we build a framework which is agnostic of the underlying sharding consensus components (i.e. intra-shard consensus, cross-shard transaction processing, and shard formation), and put into practice the ideas described in \Cref{Cobshard} and \Cref{time_frag}?

This can be done introducing another independent chain, which we call \emph{Synchronization Chain}, which is maintained by the network and has two main scopes:
\begin{itemize}
    \item \emph{synchronize the work of the nodes who work in different shards}: the Synchronization Chain dictates the beginning and the end of the time-slots and hash-links the blocks that have been legitimately created in time.
    This can be done in the following way: after every time-slot, the nodes working at the maintenance of the Synchronization Chain reach consensus on the set of blocks which have been created (by the nodes prescribed by $L$) and broadcast in time during that time-slot.
    The consensus is reached on the digest of these blocks, therefore a block of the Synchronization Chain contains the list of blocks created in time for each shard, and when this block gets published, the network starts the new time-slot and knows which blocks have passed the first validation (which is only about timing and legitimacy);
    
    \item \emph{deal with epoch reconfiguration}: the block of the Synchronization Chain created in the last time-slot of each epoch, together with the hash pointers to the blocks of the shards mentioned above, contains the parameters of the following epoch.
    The consensus protocol must specify how the parameter must be valued, and must specify how the nodes must create the list $L$.
    The nodes maintaining the Synchronization Chain, follow these rules to compute the parameters, and produce this larger block during the last time-slot of an epoch.
    With this information the network activates the next epoch and the new time-slots.
\end{itemize}

    \begin{table}[ht]
    \centering
    \begin{tabular}{|p{0.22\textwidth}| p{0.5\textwidth}}
    \cmidrule[1pt]{1-1}
    \centering \textbf{HEADER} &\\
    \cmidrule[1pt]{1-1}
    $H(S_{h,i-1})$ & hash-pointer to previous Synchronization Block\\

    \cmidrule[1pt]{1-1}
    \centering \textbf{DATA} &\\
    \cmidrule[1pt]{1-1}
    $H(B_{h,i}^s)$ &\\
    \cmidrule(lr){1-1}
    $H(B_{h,i}^1)$ & hash-pointers to the blocks created\\
    \vdots & during $i$-th time-slot of epoch $h$ on shard $s$\\
    $H(B_{h,i}^m)$\\
    \cmidrule[1pt]{1-1}
    \centering \textbf{EPOCH DATA} & (only in blocks created after the last time-slot of epoch $h$)\\
    \cmidrule[1pt]{1-1}
    $\mathtt{parameters}$ & value of parameters for following epochs\\
    \cmidrule(lr){1-1}
    \texttt{List} $L$ & list of nodes that will create blocks in the assigned shards in the next epoch\\
    \cmidrule[1pt]{1-1}
    \end{tabular}
    \caption{A block of Synchronization Chain created in a time-slot of epoch $h$.}
    \label{tab:synchroblock}
\end{table}

In this context, Cob is well-suited to be used as the consensus protocol for the Synchronization Chain.
In fact, besides being leaderless, a property which guarantees the authenticity of the data agreed upon, it efficiently carries out the consensus process in parallel on each component of the list of events, which is essential in this kind of applications, as emphasized in \Cref{protocol}.
Since the agreement on some components of the list of observed events might be reached on $\bot$, the blockchain consensus protocol must determine some default values for the epoch parameters.
That is, there should be a rule which decides the value of the parameters  to be used when agreement is reached on $\bot$.
For example, the configuration of the previous epoch could be maintained, otherwise the network could adopt some fixed configuration.
This design choice depends on the application context.

The protocol that describes the use of the Synchronization Chain can be summarized in the following way:
\begin{protocol}{Synchronization Chain}
\begin{itemize}
    \item \texttt{Epoch reconfiguration and timing evaluation}: the network executes an instance of Cob to decide the new epoch parameters, and the blocks created in time during the current time-slot.
    As soon as the network reaches agreement, the block of the Synchronization Chain is published, advertising the parameters and the digest of the blocks created in time. The new epoch begins. 
    
    Go to \texttt{Certificate creation and start of a new time-slot}.
    
    \item \texttt{Timing evaluation}: the nodes of the network start executing Cob when their own private clock signs time $t$, trying to reach consensus on the digest of the newly created blocks, one for each shard.
    
    Go to \texttt{Certificate creation and start of a new time-slot}.

    \item \texttt{Block creation}:
    the nodes in charge, according to the list $L$ published at the end of \texttt{Epoch reconfiguration and timing evaluation}, build a block of transactions and before time $t-2\lambda$ broadcast their block.
    
    If the current time-slot is the last of the epoch, go to \texttt{Epoch reconfiguration and timing evaluation}, otherwise go to \texttt{Timing evaluation}.
    
    \item \texttt{Certificate creation and start of a new time-slot}:
    as soon as a player receives a certificate for the newly created blocks (produced by the network using Cob), it marks the end of the current time-slot, and the beginning of the new time-slot.
    Therefore the nodes can reset their same-speed clocks (which will be delayed by at most by the diffusion time of a certificate, namely $\lambda$) and return to \texttt{Block creation}.
\end{itemize}
\end{protocol}

Since Cob is a consensus protocol with strong consistency, the information included in the blocks of the Synchronization Chain are final.
Moreover, thanks to the fact that Cob is leaderless, the evaluated fields are trustworthy.
Therefore, the Synchronization Chain can be seen as a trusted third party who communicates the outcome configuration for the following epoch based on the information the network has collected during the current (but possibly also previous epochs).

The intra-shard consensus protocol can be weak, which allows lower communication consumption, however, due to to the Synchronization Chain's timing evaluation, there is a strong consistency consensus on the legitimately created blocks.
This simplifies Cross-shard transaction processing, since the nodes working in different shards know which transactions can potentially become final.

\section{Performance analysis}
\label{performance}

We can now take the performance analysis of Cob included in \cite{Cob} and apply it to the use case discussed here.
The analysis is focused on comparing the amount of data broadcast in the network during a instance of Cob executed on a list with $\ell$ components with the amount of data broadcast in an execution of $\ell$ instances of Algorand to reach Consensus on each relevant information regarding the observed events.

In order to provide a comparison which fits the use case of the Synchronization Chain described in \Cref{synchrochain}, we must identify a reasonable number of parameters necessary to perform the epoch reconfiguration, then we can vary the number of shards and time-slots to determine the number of elements in the list $L$.

Once this is done, we can quantify the weight of messages broadcast at the end of the last time-slot of each epoch, when the epoch reconfiguration is performed, and the weight of messages of all the other time-slots, when the network must notify only the blocks created within the end of the current time-slot.
For sake of simplicity we will assume the number of time-slots during an epoch is fixed, so that we can vary only the number of shards.

The number of components $\mathit{Nc}_{e}$ that must be agreed upon in the last stime-slot of epoch $e$ can be parameterized as follows:  $\mathit{Nc}_{e}=\alpha + \beta \mathit{Ns}_{e+1}+\mathit{Ns}_{e}$ where $\mathit{Ns}_x$ is the number of shards in epoch $x$, $\alpha$ is the number of parameters defining the general configuration of the platform, and $\beta$ is the number of parameters specifying some properties of each shard of the next epoch (e.g. the nodes designated to work on a given time-slot). 

Let us fix the number of time-slots for epoch to 10, and choose the parameters $\alpha=20$ and $\beta=10+1$ (these values are chosen in the same order of magnitude as the ones relative to the blockchain Quadrans \cite{battagliola2021quadrans}, where 10 of the $\beta$ parameters characterizing the shards are the nodes assigned to a given time-slot, for each shard).
For what concerns the regular time-slot, the number of components that must be agreed upon is simply $\mathit{Ns}_e$, i.e. the number of blocks that should be created (note that this number is considered also in the last time-slot).

\begin{figure}
\begin{minipage}{0.48\textwidth}
        \centering
        \includegraphics[width=\textwidth]{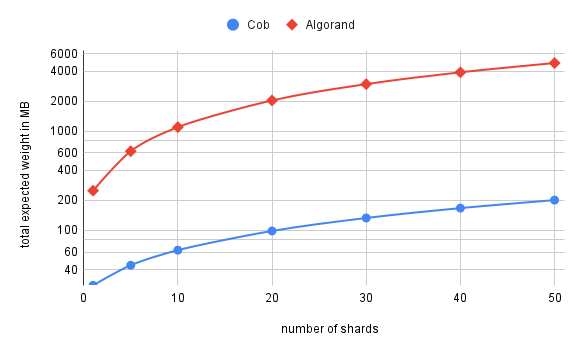}
        \caption{Amount of data broadcast in the network (in MB) using Algorand or Cob during the last time-slot of each epoch, in terms of the number of shards. Linear scale in the x-axis and logarithmic scale in the y-axis.}
        \label{epochblock}
    \end{minipage}\hfill
    \begin{minipage}{0.49\textwidth}
        \centering
        \includegraphics[width=\textwidth]{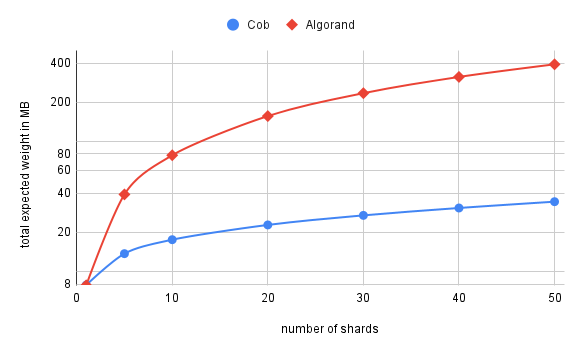}
        \caption{Amount of data broadcast in the network (in MB) using Algorand or Cob during any time slot of each epoch but the last one, in terms of the number of shards. Linear scale in the x-axis and logarithmic scale in the y-axis.}
        \label{synchroblock}
    \end{minipage}
\end{figure}

The results of the comparison are presented in \Cref{epochblock} and \Cref{synchroblock}, and are based on the performance analysis included in \cite{Cob} for the values of parameters $\alpha$ and $\beta$ mentioned above. For sake of simplicity we also assumed $\mathit{Ns}_{e}=\mathit{Ns}_{e+1}$, i.e the number of components in the last time-slot is $20+ 12 \mathit{Ns}_{e+1}$.

\section{Conclusions}
\label{conclusions}
In this paper it is shown how the consensus protocol Cob, presented in \cite{flamini2022multidimensional, Cob}, can be useful for designing sustainable sharding-based consensus protocols for blockchains, as suggested in the original papers \cite{flamini2022multidimensional,Cob}.
The key concept is the following: in an architecture that pre-assigns time-slots to nodes, the node assigned to a given time-slot in a shard is common knowledge, and the network is in agreement about this.
The same holds for the quality evaluation of the transactions included in the blocks: every honest node can determine whether a given block contains valid transactions according to the chain of blocks to which it is connected.
In fact, the consensus protocol must define how the ledger can evolve and, given a static status of the ledger, which transactions can be appended.
The only thing which remains subjective for each node is the moment in which a message is received.
Someone might have received it in time, someone else might have received it late.
These messages may be blocks of transactions or data useful for the epoch reconfiguration, anyways, it is essential for the network to have a clear image of the status of the evolving system (the blockchain), in particular when the system is maintained by several groups working in parallel, which is the case of a blockchain that uses sharding to scale.
We propose a solution to this problem using Cob, so that consensus can be reached on these subjective data (the network decides on the basis of what the majority of the nodes have observed) and every node in the network can have the same view on how the ledger is evolving.

\section*{Declarations}
The authors have no relevant financial or non-financial interests to disclose.
The authors have no conflicts of interest to declare that are relevant to the content of this article.
The authors are members of the INdAM Research group GNSAGA.
The first author acknowledges support from Eustema S.p.A. through the PhD scholarship.
The authors have no financial or proprietary interests in any material discussed in this article.

\bibliographystyle{plain}
\bibliography{Bibliography}
\clearpage
\end{document}